\documentclass[aip,apl,preprint,preprint,superscriptaddress,amssymb]{revtex4-1}
\usepackage{graphicx}
\usepackage[latin1]{inputenc}

\begin{document}

\title{Tunable magnetic states in h-BN sheets}

\author{Eduardo Machado-Charry} 
%\email{eduardo.machado-charry@cea.fr}
\affiliation{Nanosciences Fondation, 23 rue des martyrs, 38000 Grenoble, France}
\affiliation{Laboratoire de simulation atomistique (L\_Sim), SP2M,INAC,CEA-UJF, Grenoble, F-38054, France} 

\author{Paul Boulanger}
\affiliation{Laboratoire de simulation atomistique (L\_Sim), SP2M,INAC,CEA-UJF, Grenoble, F-38054, France} 

\author{Luigi Genovese}
\affiliation{Laboratoire de simulation atomistique (L\_Sim), SP2M,INAC,CEA-UJF, Grenoble, F-38054, France} 

\author{Normand Mousseau}
%\email{normand.mousseau@umontreal.ca}
\affiliation{D\'epartement de Physique and RQMP, Universit\'e de Montr\'eal, C.P. 6128, Succursale Centre-Ville, Montr\'eal (QC) Canada H3C 3J7}

\author{Pascal Pochet}
%\email{pascal.pochet@cea.fr}
\affiliation{Laboratoire de simulation atomistique (L\_Sim), SP2M,INAC,CEA-UJF, Grenoble, F-38054, France} 

\date{\today}

\begin{abstract}

Magnetism in 2D atomic sheets has attracted considerable interest as its existence could allow the development of electronic and spintronic devices. 
The existence of magnetism is not sufficient for devices, however, as states must be addressable and modifiable through the application of an external drive. We show that defects in hexagonal boron nitride present a strong interplay between the the N-N distance in the edge and the magnetic moments of the defects. By stress-induced geometry modifications, we change the ground state magnetic moment of the defects. This control is made possible by the triangular shape of the defects as well as the strong spin localisation in the magnetic state.

\end{abstract}

%\pacs
%{
%75.75.-c % Magnetic properties of nanostructures, Quantum dots, Quantum wells
%73.22.-f % Electronic structure of Nanocrystalline materials, Nanoparticles, Nanotubes, solid clusters, condensed matter nanoscale materials
%64.70.Nd % Nanoscale materials: structural transitions in 
%61.72.jd % Vacancies, in crystals
%}

\maketitle

Studied for only a few years, two-dimensional atomic sheets have shown a remarkably rich set of properties that could play a significant role in the development of spintronics and of the next generation of electronic devices. In particular, hexagonal boron nitride (h-BN) sheets have gained significant interest recently due to their similarities with graphene~\cite{Dean2010,doi:10.1021/nl2011142}, both being 2D honeycomb lattices. However, h-BN differs from the latter on a number of properties. For example, it presents a wide bandgap associated with the partially ionic character of its bonding in contrast with the semi-metallic character of graphene~\cite{PhysRevB.79.115442}. As BN materials are generally more resistant to oxidation than graphene, h-BN should also be more suitable for applications at high-temperatures~\cite{Liu2007}. Its binary composition also increases the possibility for creating and positioning defects with specific properties, opening the door to a very pointed design and a rapid incorporation into technological devices. Recent studies, for example, have revealed interesting features of defects created by electron beam irradiation of h-BN sheets~\cite{ISI:000266207700046,Meyer2009a,Kotakoski2010,Alem2009}. High-resolution transmission electron microscopy (HRTEM) shows triangular multivacancy structures with discrete sizes but with a unique orientation. By means of the exit-wave reconstruction method in combination with HRTEM images, Jin \textit{et. al.}~\cite{ISI:000266207700046} were able to assign a chemical label to the atoms surrounding the defects and demonstrate that these triangular multivacancies have N-terminated zigzag edges. This feature suggests that the boron atoms are easier to knock out than nitrogen under electron irradiation. This bias could facilitate the manufacturing of functional devices with well-controlled defect structures and fully justifies a more careful study of these defects' properties. Du \textit{et al.}~\cite{Du2009} have showed that these triangular vacancies exhibit a total magnetic moment proportional to the number of nitrogen dangling bonds in the N-terminated edges. This result is in line with calculations that predict large magnetic moments for BN nanoribbons with a ferromagnetic orientation for the N-zigzag edge and an antiferromagnetic for the B-edge~\cite{Huang2011a}. While these numerical results have not yet received an experimental confirmation, they are supported by a characterization of defect-induced paramagnetism in graphene that demonstrates the validity of similar calculations in 2D carbon~\cite{ISI:000300929800013}. It is therefore tempting to see applications of h-BN in the current electronic industry but also for spintronic devices. For this purpose, however, static magnetic moments are not really sufficient, and tunable spin states remain the target. 

In this letter, we focus on the magnetic properties of multivacancy holes in h-BN sheets and their relation to structural features of the holes. Using spin-polarised density functional (DFT) calculations as implemente in BigDFT~\cite{ISI:000257468100014}%,genovese:054704}
, we have characterized in detail the electronic state of the small multivacancy defects and explored approaches to tune their spin state. We have focused more particularly on the destabilisation of the reconstructed corner under tensile strain. Our results show that, under a given uniaxial deformation, the atoms surrounding a specific type of triangular multivacancy defect can undergo a pronounced geometric reconstruction. This reconstruction strongly affects the local electronic states, modifying the defect's total magnetic moment. These results reveal a tight interplay between the local strain and the local spin state, as the electronic states remain strongly localised during the transformation. This opens new ways to exploit the magnetic properties of defects in BN sheets.

Calculations were done in a cell size that guarantees negligible elastic interactions between the periodic defects. To this end we use an orthorhombic
supercell ($35.2 \times 34.8$~\AA$^2$) with 448 atoms and only the $\Gamma$ point. In the perpendicular direction we use free boundary conditions, i.e. infinite vacuum. The wavelet grid is chosen such that the energy per atom is accurate within 5 meV, which corresponds to a grid spacing of 0.35 bohrs and a localisation radius of 9.6 bohrs. Geometries are considered optimized when the forces on atoms are less than $15~\mathrm{meV}/\text{\AA}$. We use a PBE exchange-correlation functional together with HGH pseudopotentials~\cite{Krack-2005}.

\begin{figure}
 \includegraphics[width=0.35\textwidth]{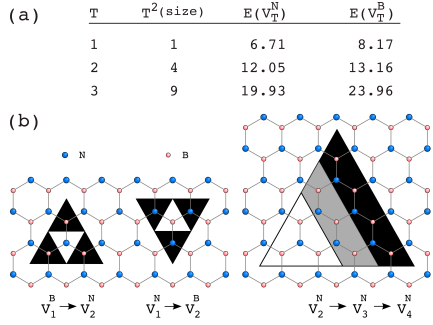}
 \caption{(Color online) (a) Formation energies (in eV) for different N-edge and B-edge terminated vacancy sizes $T^2~(T=1,2,3)$. The values are given per defect. (b) Growth route for $V^N_2$, $V^B_2$ and N-terminated bigger holes. B atoms are in red and N atoms in blue.}
 \label{fig1}
 \end{figure}

As starting point we address the formation energy of multivacancies in h-BN. This calculation requires the use of a suitable chemical potential $\mu_N(\mu_B)$ for N(B) atoms. Because of the wide bonding diversity for both elements, we select, as point of comparison, compounds with an $sp^2$ environment similar to the h-BN sheet. Thus, knowing the chemical potential $\mu_C$ in graphene, $\mu_N$ is obtained with the help of a sheet of g-C$_3$N$_4$~\cite{ISI:000259913900015}, which is a crystalline graphitic carbon nitride compound. Hence, the formation energy for a vacancy defect can be estimated from the expression:
\begin{equation}
E_{D_T}=E^{X-D_T} -( E^X - n_B\mu_B -n_N\mu_N ),
\end{equation}
where $E^{X-D_T}$ is the total energy of the h-BN sheet with a defect $D_T$ and $E^X$ is the total energy of the perfect system. In the following we consider several triangular multivacancies defects $V_T$, where $n_B$ and $n_N$ number of B and N atoms have been removed, such as $T=\sqrt{n_B+n_N}$. The correponding defects of size $T^2$ are  either with N ($V^N_T$) or with B  ($V^B_T$) edges. Calculated formation energies for $T=1,2,3$ are displayed in Fig. \ref{fig1}(a). $V^N_1$ is found to be more favourable than $V^B_1$. However, the incident electron energy in the experiments~\cite{ISI:000266207700046,Meyer2009a} is rather large with respect to the emission energy threshold for both atoms~\cite{PhysRevB.75.245402}. Thus, it is likely that both $V^N_1$ and $V^B_1$ are formed during the early irradiation time, while the former is the dominant one. Our results explain this discrepancy and offer a growth route for bigger defects. The direct formation of the $V^N_2$ defect ($12.05$ eV) is unfavoured with respect to an indirect formation starting from $V^B_1$ ($+3.88$ eV), see Fig. \ref{fig1}(b). This would explain the observed low concentration of $V^B_1$ in favor of  $V^N_2$. Conversely, the formation of the $V^B_2$ defect is unlikely  from both routes, direct ($\sim13$ eV) and indirect, $V^N_1 \rightarrow V^B_2$ ($\sim6.5$ eV). This could explain the large presence of $V^N_1$ in experimental observations rather than $V^B_2$. The next step, the formation of $V^N_3$, is done by removing one biline  (Fig. \ref{fig1}(b)) from one edge of $V^N_2$ for a cost of only $7.9$ eV. The further transformation, $V^N_T \rightarrow V^N_{(T+1)}$, is then expected to occur with an even lower energy cost. This proposed  growth route is in agreement with the main experimental observations: a low concentration of $V^B_1$, a high concentration of $V^N_1$ and N-terminated big holes. The validity of this hypothesis could be easily assessed by varying the irradiation time and following the evolution of the defects. Of particular importance is the fact that by tuning the irradiation time it would be possible to restrict the defect family to $V^N_1$, $V^B_1$ and $V^N_2$, while long irradiation time always lead to large holes~\cite{ISI:000266207700046,PSSR:PSSR201105262}.

%figure 1
 
\begin{figure*}
 \includegraphics[width=0.95\textwidth]{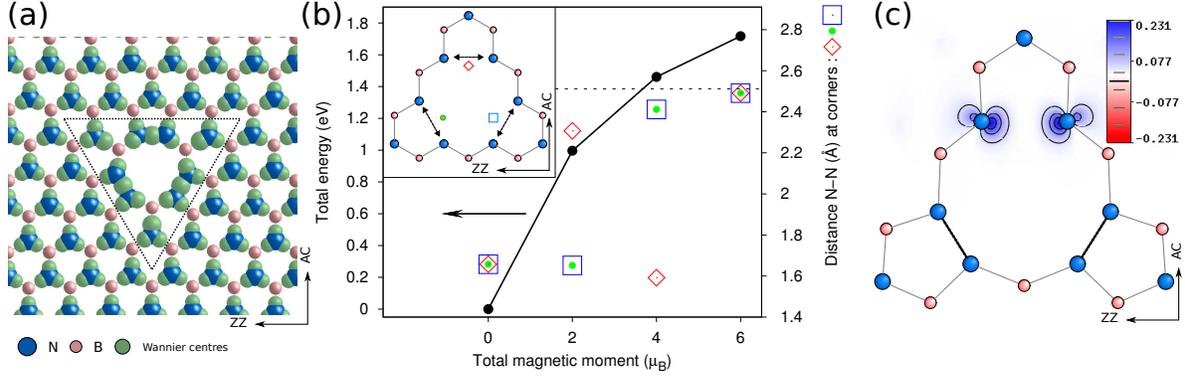}%
 \caption{(Color online). (a) Optimized geometry and Wannier centers of a h-BN sheet with one $V^N_2$ defect. B atoms are in red, N atoms are in blue and Wannier centers are in green. Zigzag (ZZ) and armchair (AC) directions are shown. The Wannier states outside the triangle are not disturbed by the defect. (b) Total energy and N-N distance at each corner (inset) as a function of the imposed magnetic moment. (c) Optimized geometry for the system with $M=2~\mu_B$ along with the difference in the spin magnetic charge density $(\rho_\uparrow -\rho_\downarrow)$ projected on the plane of the monolayer. For clearness, only atoms along the borders of the vacancy are included. \label{fig2}}
 \end{figure*}

The calculated total magnetic moment for $V^N_1, V^N_2$ and $V^N_3$ defects is 1, 0 and 3~$\mu_B$, respectively, and doesn't follow Lieb's theorem~\cite{Lieb1989} contrary to graphene~\cite{PhysRevB.84.214404}. This violation is mainly due to strong reconstructions. Indeed, calculations show that the $V^N_T$ defects, with $T>1$, undergo a significant reconstruction with the formation of a N-N bond in each corner of the triangle. Figure~\ref{fig2}(a) shows the ground state geometry of a $V^N_2$ defect along with the Wannier centers from a Wannier projection~\cite{Wannier90} of the electronic states. The N-atoms at the corners form pairs with an average distance of only 1.66~\AA, 0.85~\AA~shorter than in the perfect h-BN sheet. This pairing satisfies all dangling bonds, producing a homopolar $\sigma$ bond between the two N atoms which can be seen by the presence of only one Wannier center in the middle of the pair. This bonding reduces the total energy as well as the total magnetic moment associated with the hole to zero. The electronic states modified by the defect are very localised on the N atoms surrounding it and their closest neighbors (in the triangle in Fig.~\ref{fig2}(a)), suggesting a little communication between nearby holes.
A similar behavior is seen on the $V^N_3$ defect. Leaving an unpaired electron on each edge of the triangle, $V^N_3$ displays a net total magnetic moment of $3~\mu_B$.  By extension, for any triangular defect, the magnetic moment at ground state, $M_{GS}$, is held by dangling bonds and should go as $M_{GS}=3\times T-6$ for $T>1$. Thus, the electronic and magnetic properties of h-BN sheets are strongly affected by the size of the vacancy defects.

% figure 2

Although the ground state of a V$^N_2$ defect is clearly $M=0~\mu_B$, it is interesting to study the impact of non-zero total magnetic moments both on its geometry and its spin localisation. To do so, we did spin-constrained calculations by fixing the total magnetic moment on the whole system and minimizing the total energy, allowing full geometrical relaxation, in order to identify the best ordering of the local spin polarization. Figure \ref{fig2}(b) shows the relative total energy and the N-N distance at each corner (symbols defined in the inset) as a function of the imposed total magnetic moment --- $M=0$, 2, 4 and $6~\mu_B$. As the total magnetic moment goes up, the total energy of the system increases monotonically as the homopolar N-N bonds are broken. With $M=2~\mu_B$, the lowest-energy state with a non-zero magnetic moment, we observe a symmetry breaking.
Two N atoms in one corner move away from each other, up to 2.3~\AA, while atoms in the two remaining pairs move slightly closer. Figure \ref{fig2}(c) shows the optimized geometry of the atoms around this defect with $M=2~\mu_B$ and the difference in the charge density $(\rho_\uparrow-\rho_\downarrow)$ projected on the plane of the monolayer, it shows that the extra spins arrange themselves predominantly in the two dangling N bonds arising by the opening of one corner. As the total magnetic moment is further increased (Fig. \ref{fig2}(b)) the remaining two pairs are broken in sequence, bringing the interatomic distance between corner-sharing N to the lattice value for second neighbor N atoms, i.e., 2.51~\AA. As the triangular defect size $T^2$ goes up, so does the number of unpaired electrons on the edges, but the N-pairing at the three corners remains. Thus, to ensure spin localisation the total magnetic moment that can be imposed on these triangular defects is $M=M_{GS} + 2~(+4,+6)$, irrespective of the size.

If a change in the total magnetic moment can affect the geometrical reconstruction, will a structural deformation impact the total magnetic moment? In other words, is it possible to modify the magnetic moment of a defect by imposing a structural deformation?  Given that increasing $M$ induces symmetry breaking, it is natural to first look at the application of an uniaxial tensile strain, either along the armchair or the zigzag direction, that would have the same impact. From the results on the effect of $M$ on the reconstruction at the edge of the triangular defect (Fig. \ref{fig2}(b)), we can expect that a tensile strain in the zigzag direction would lead first to the rupture of the upper corner of the triangle since the N-N bond in this corner is oriented in the zigzag direction. Such a deformation would imply a transition from $M_{GS}\rightarrow +2~\mu_B$. Similarly, an uniaxial strain in the armchair direction would affect first the two N-N bonds in the lower corners of the triangle, with an expected transition from $M_{GS}\rightarrow +4~\mu_B$.

\begin{figure*}
 \includegraphics[width=0.95\textwidth]{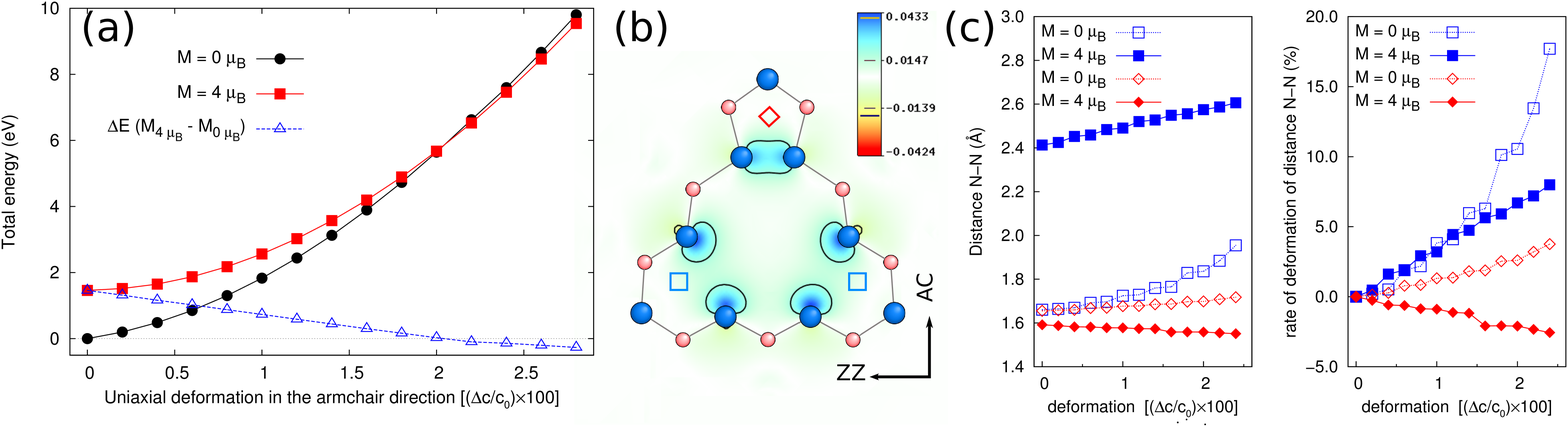}%
 \caption{(Color online). (a) Evolution of the total energy for the system with a $V^N_2$ defect as a function of the uniaxial deformation in the armchair direction. Open triangles, energy difference, $\Delta E$, between magnetic states $M=4~\mu_B$ and $M=0~\mu_B$. (b) Optimized geometry and the Wannier states projected on the plane of the monolayer for the system with $M=4~\mu B$ at an uniaxial deformation in the armchair direction of 2.4 \%. For clearness, only atoms along the borders of the vacancy are included. (c)  Absolute and relative evolution of the N-N distance in a system with a $V^N_2$ defect as a function of uniaxial deformation in the armchair direction. Diamond symbols refer to the N-N bond in the zigzag direction, while square symbols are the mean value of both N-N distances as they are presented in (b).\label{fig3}}
\end{figure*}

While the relation between an external strain and magnetic state appears relatively clear, and it was shown, recently, that strain can indeed enhance magnetic vacancies~\cite{kan:072401}, it is the evolution of the total energy for the various magnetic states as a function of the deformation that determines whether these transformations are physically relevant or not. Figure~\ref{fig3}(a) shows the evolution of the total energy as a function of the uniaxial deformation in the armchair direction for the system with a $V^N_2$ defect in two definite magnetic states $M=0~\mu_B$ and $M=4~\mu_B$. For clarity, we also plot the energy difference, $\Delta E$, between both magnetic states. At zero~\% of uniaxial deformation, as calculated before, the ground state with $M=0~\mu_B$ is 1.46 eV more stable than that at $M=4~\mu_B$. This difference reduces monotonically, almost linearly, as the deformation increases, showing a crossing at a deformation of about 2.0\%, at which point the $M=4~\mu_B$ becomes more stable than the one with zero magnetic moment (Fig.~\ref{fig3}(a), $\Delta E$ line) by as much as 0.2 eV at 2.4\% of deformation.

 % figure 3

Figure~\ref{fig3}(b) shows the optimized geometry for a system with a total magnetic moment of $4~\mu_B$ at 2.4 \% of deformation in the armchair direction as well as the projections of Wannier states on the plane of the monolayer. As can be seen, the upper bond, which is perpendicular to the deformation, remains closed while there is bond breaking in the bottom corners. Since all other atoms maintain their original coordination, the spin magnetic charge density also remains localised precisely on those unpaired dangling bonds.
This transition occurs because a large fraction of the strain is released, anharmonically, by the stretching of the N-N distance in the two lower corners of the triangular defect (most parallel to the deformation) as seen in Fig. \ref{fig3}(c). These extend by about 10 \% at a 2.0 \% of deformation for both states, stretching the N-N bonds in the $M=0~\mu_B$ case to 1.8~\AA. The transition from $M_{GS}\rightarrow +4~\mu_B$ in the V$^N_2$ defect with an uniaxial strain applied along the armchair direction is therefore due to a very subtle balance between the overall elastic relaxation around the defect as the N-N bonds parallel to the deformation are broken and the cost of placing two spins nearby. For this mechanism, the direction of the deformation is important, for example in the zigzag direction the transition occurs at a greater deformation. Moreover, it also depends of the size of the defect: it is impossible to reproduce this phenomenon with uniaxial deformation along both the armchair and zigzag direction in the larger V$^N_3$ defect (not shown).
In those cases, the cost of the elastic deformations of the whole system is higher than the energy gain obtained from adding pairs of spins at the corners of the triangular defects. However, since the arising magnetic moment remains localised, a deformation of the whole system around the defect is in principle not needed. The stress-induced deformations presented here clearly show that the N-N distance and the magnetic moment of the defect are strongly related to each other. It is important to point out that also a different mechanism which would only constrain the N-N bondings \emph{close to the defect} would do the job. For example, this effect could also be used for molecular detection, where a molecule fixing itself on a defect, of various size, could induce a large enough structural deformation to induce a change in the spin state. This bias for N-terminated defects together with a well controlled irradiation time could facilitate the creation of arrays with V$^N_2$ defects, opening the door of a tunable spin state nanostructure.

%\begin{acknowledgments}
We acknowledge support by the Nanosciences Fondation (MUSCADE), ANR (NANOSIM-GRAPHENE), NSERCC, the FQRNT and the CRC Foundation. This work was performed using HPC resources from GENCI (Grant 6323), Tera-100 from CEA-DAM, and Calcul Qu\'ebec.
%\end{acknowledgments}

%Merlin.mbs v4.21 2009-07-09.
%

\end{document}